\documentclass{article}
\usepackage[T1]{fontenc} % add special characters (e.g., umlaute)
\usepackage[utf8]{inputenc} % set utf-8 as default input encoding
\usepackage{ismir,amsmath,amssymb,cite,url}
\usepackage{balance}
\usepackage{graphicx}
\usepackage{color}
\usepackage{colortbl}
\usepackage{booktabs}
\usepackage{subfig}
\usepackage[table]{xcolor}
\title{Surprising Patterns in Musical Influence Networks}

\threeauthors
  {Flavio Figueiredo} {UFMG \\ {\tt flavio@dcc.ufmg.br}}
  {Tales Panoutsos} {UFMG \\ {\tt tales.panoutsos@dcc.ufmg.br}}
  {Nazareno Andrade} {Enveritas \\ {\tt nazareno@gmail.com}}

\begin{document}

\maketitle

\begin{abstract}
Analyzing musical influence networks, such as those formed by artist influence or sampling, has provided valuable insights into contemporary Western music. Here, computational methods like centrality rankings help identify influential artists. However, little attention has been given to how influence changes over time. In this paper, we apply Bayesian Surprise to track the evolution of musical influence networks. Using two networks—one of artist influence and another of covers, remixes, and samples—our results reveal significant periods of change in network structure. Additionally, we demonstrate that Bayesian Surprise is a flexible framework for testing various hypotheses on network evolution with real-world data.
\end{abstract}

\section{Introduction} \label{sec:intro}

In recent years, the study of how artists influence~\cite{shalit2013modeling,figueiredo2019disruption}, collaborate~\cite{park2007social,teitelbaum2008community,bae2016scale,andrade2016exploring,gunaratna2011mpb,gleizer2003jazz,uzzi2005collab,silva2004complex} and sample~\cite{bryan2011musical,van2012sample} one another has lead to valuable knowledge on the evolution of contemporary western music\footnote{A limitation primarily due to available data. However, some exceptions exist such as:~\cite{silva2004complex,uzzi2005collab,gunaratna2011mpb}}. A large fraction of these studies rely on social network analysis. Music data is studied via the relationships between songs, albums, or artists in these networks, or {\em musical influence networks}.

When we look at influence networks evolving (e.g., considering the rise of new artists and changes in network structure), one question arises: considering the evolution of the network, {\it what was surprising}, as time passed. Take as one example The Beatles. Saying The Beatles are highly influential nowadays is relatively easy. However, understanding The Beatles early in their career may be more complex. Here, we propose both a formal definition of Bayesian Surprise~\cite{itti2009bayesian,itti2005principled,baldi2010bits} for rankings in complex networks. We aim to answer what is (or was) surprising as the musical influence network grew.

To perform our study, we explore two human-curated datasets of music influence. One from the AllMusic Guide\footnote{\url{https://www.allmusic.com/}} and another from WhoSampled\footnote{\url{https://www.whosampled.com/}}. Our study uses both datasets via temporal views of Pagerank~\cite{page1999pagerank} and Disruption~\cite{funk2016dynamic} centrality scores.

Bayesian Surprise allows the testing of multiple hypotheses over a time-evolving dataset. We are the first to show that it provides a principled approach to combine both centrality scores and multiple hypotheses in a single measure for music influence -- or even social -- networks. Combining different centralities allows for a more expressive analysis, as each has its interpretation. Pagerank may be seen as an overall influence on an artist. In contrast, even non-influential artists may be disruptive\footnote{Disruption focuses on an artist's capacity of aggregating influence on itself diverging influence from neighboring nodes.}~\cite{figueiredo2019disruption}. The framework allows us to combine different centralities and hypotheses to unveil interesting trajectories of influence. 

\section{Related Work} \label{sec:rw}

In the past, several authors have looked into musical influence networks. These efforts ranged from looking at Classical music~\cite{bae2016scale}, to Jazz~\cite{andrade2016exploring,gleizer2003jazz}, Broadway Songs~\cite{uzzi2005collab}, as well as Popular Brazilian Music~\cite{silva2004complex,gunaratna2011mpb,andrade2020measuring}. With regards to larger and more general datasets, the AllMusic guide and the WhoSampled website have also been the focus of several prior endeavors~\cite{figueiredo2019disruption,serra2012measuring,morton2015acoustic,atherton2016said,shalit2013modeling}.

In these previous efforts, musical influence is captured by a graph, modeling the network. Here, nodes are either artists or songs. Edges encode the influence (e.g., sampled, was influenced by, or collaborated with). With these graphs, social network analysis~\cite{easley2010} analysis plays a pivotal role in understanding the importance of music.% as a cultural production.

Most of these analyses, however, fail to capture temporal aspects of the influence network. With this regard, Andrade and Figueiredo~\cite{andrade2016exploring} looked into the latent structure of Jazz collaborations via Markovian models. Even though temporal ordering is necessary for Markovian models, the authors do not look into social network properties over time. 
Here, Shalit, Weinshall, and Chechik~\cite{shalit2013modeling} consider time in the evaluation for a method to predict interest, but no insights into network evolution are provided. Other authors similarly limit their temporal analyses to overall descriptive statistics per year~\cite{bryan2011musical,figueiredo2019disruption,andrade2020measuring}.

This is the first work to adapt Bayesian Surprise as a tool that provides insights into network evolution. The measure can pinpoint each node (artist) point in time where the topology of the network influenced the artist's centrality. Next, we present the background required to understand our methodology.

\section{Networks and Surprise}

We now discuss our notation and ranking scores for the musical influence networks, Section~\ref{sec:min}. We also take the time to explain Bayesian Surprise and how we adapt it for rankings in Section~\ref{subsec:bayes}.

\subsection{Influence Networks and Centrality} \label{sec:min}

Let $\mathcal{G}_t = \{\mathcal{V}_t, \mathcal{E}_t\}$ be a weighted directed graph. $\mathcal{V}_t$ captures the set of nodes, whereas $\mathcal{E}_t$ captures edges. For both datasets this graph captures influences across artists. The subscript $t$ here indicates that our graph is time evolving in discrete steps (decades or years). In other words, $t$ defines the snapshot time of our graph. $\mathcal{G}_{2010}$, is the graph of influences formed up to and including 2010.

We assume nodes are enumerated, i.e., $n_i \in \mathcal{V}_t$ determines a node $n_i$ with each node defining an artist. Edges are similarly defined, $e_k \in \mathcal{E}_t$ with $e_k = (n_i, n_j)$ representing an edge from node $n_i$ to node $n_j$. Our graphs are directed, thus i.e., $(n_i, n_j) \neq (n_j, n_i)$. As we are studying musical influence across artists, we impose that $j \neq i$, stating that an artist may not influence itself\footnote{In one of our datasets, the WhoSampled data, artists may sample and cover their own work, and thus we remove this characteristic from graph.}. Both node and edge ids are maintained in subsequent snapshots. We also define edge weights, $weight(\mathcal{G}_{t}, e_k)$, the number of times one artist mentions the other as an influence. Finally, we note that our graphs evolve cumulatively, i.e., $weight(\mathcal{G}_{t'}, e_k) \geq weight(\mathcal{G}_{t}, e_k)$, when $t' > t$. At each time-step $t$, nodes are ranked according to Pagerank~\cite{page1999pagerank} and Disruption~\cite{funk2016dynamic} centrality. That is, each node has a score $s_{t,i} = score(\mathcal{G}_{t}, n_i)$. 

The first score, Pagerank~\cite{page1999pagerank}, outputs a probability distribution over the entire set of nodes in the network. If we consider a random walker on the graph, this probability distribution may be interpreted as the chance a random walker ``lands'' on node $n_i$ after sufficient steps. The walker jumps from node $n_i$ to $n_j$ with a chance proportional to $weight(\mathcal{G}_{t}, e_k)$, where $e_k = (n_i, n_j)$. 

Disruption~\cite{funk2016dynamic}, focuses on the capability of a node on concentrating influence. That is, considering that artists cite past influences, when some focal artist node, $n_a$, is mostly cited by-itself it is deemed are more disruptive. That is, if we consider that for $n_a$ there are some nodes that reference $a$'s work and at least one of $a$'s own influences. Some other nodes cite only $a$. When $a$ is cited in isolation, disruption increases. When it is not, it decreases. The final score is simply the difference between the fraction of nodes in each group. Positive numbers indicate more disruption.

%Before continuing, we point out that both scores are computed at each snapshot. %We now describe surprise for rankings.

%. For this node, there are $J\geq0$ other nodes that reference $a$'s work and at least one of its neighbors (nodes referenced by $a$). These $J$ artists may be interpreted as those who are both influenced by $a$, as well as $a$'s prior influences. Also, consider that there exists $I\geq0$ nodes reference $a$ but none of its neighbors. These nodes are influenced by $a$ only.  When $I > J$, node $a$ will concentrate influence. Thus, the node is disruptive. $I<J$, $a$'s neighbors are influence by both $a$ and it's influences, indicating that $a$ by itself may not be so important. If we consider that there are $K\geq0$ other nodes which only reference $a$'s neighbors, Disruption is:
%\begin{equation}\label{eqn:D}
%D = \frac{I - J}{I + J + K} 
%\end{equation}
%\noindent From the equation above, we can see that different from Pagerank, Disruption is not weighted.

\subsection{Bayesian Surprise} \label{subsec:bayes}

The basis of our analysis is the measure of Bayesian Surprise~\cite{itti2005principled,itti2009bayesian,baldi2010bits}, that we now explain. %(Section~\ref{subsub:inf}). %As this section diverges from the music domain for a while, readers already familiar with the subject may want to skip directly to our discussion of surprise in musical influence networks (Section~\ref{subsub:musicsup}).
%\subsubsection{Priors, Likelihoods and Posteriors} \label{subsub:inf}
Consider an arbitrary dataset $\mathcal{D}$ to understand Bayesian Surprise. Now, let us assume that data points come from some probability distribution $x_i \sim Dist(\theta)$, with $\theta$ being the parameters of the probability density function (pdf), i.e., $p(x\mid\theta)$.

A major part of statistics is focused on finding the best parameters $\theta$ for an overall model (pdf) and a dataset. Assuming that data points are independent, this comes from maximum likelihood estimation (MLE) is achieved by maximizing: 
$\theta=\arg\,max_{\theta'}\prod_{i=1}^{|\mathcal{D}|} p(x\mid\theta')$. %In other words, find parameters $\theta$ which make the data more probable.% assuming $p(x\mid\theta)$. As an example, for the Normal distribution this ultimately boils down to the sample mean $\hat{x} = mean(\mathcal{D})$ and the sample variance $s^2 = var(\mathcal{D})$, i.e., $\theta=\{\bar{x}, s^2\}$.
Bayesian Statistics provides an approach to tackle such estimation problem via the usage of 
%We are still interested in some credibility over the estimated parameters $\theta$. 
%In order estimate parameters, Bayesian Statistics employs 
prior distributions over parameters: $p(\theta)$. 
%This is what is called the prior distribution (sometimes it is also called the Hypothesis). 
%The prior is simply another model. However, in this case it is a model over parameters. We could state that $\mu \sim Normal(0, 1)$, indicating that the mean comes from a Normal distribution. We could also state that the variance is somewhere between ten and twenty, $\sigma \sim Uniform(10, 20)$. This is why this is called a prior, or Hypothesis, these are our prior beliefs. 
By employing Bayes Theorem, we are able to derive another pdf over the parameter space:
\begin{align}
    p(\theta \mid \mathcal{D}) = \frac{p(\mathcal{D} \mid \theta) p(\theta)}{p(\mathcal{D})} = \frac{p(\mathcal{D} \mid \theta) p(\theta)}{\int_\theta p(\mathcal{D} \mid \theta) p(\theta) d\theta}
\end{align}
Here, $p(\theta \mid \mathcal{D})$ is called the posterior. It states, after observing the data, what is the probability distribution (which captures uncertainty) over the parameters. %

%Even though it may be awkward to think about this probability distribution, it may seem at a first glance unrelated to the dataset as it talks parameters, notice that this distribution is still a byproduct of the likelihood (it appears on the numerator). Thus the likelihood, or the functional form of how the parameters capture the data, will still play a large role on the posterior. This role is however weighted by the prior beliefs, i.e., the numerator is $p(\mathcal{D}, \theta) = p(\mathcal{D} \mid \theta) p(\theta)$. In the end, the dataset will still affect the posterior, thus regardless of the functional form of the prior.

%Different from the MLE approach, we have an estimate of the pdf of parameters. MLE commonly explores point estimates only for each parameter. Such distributions are the basis for Bayesian Surprise~\cite{itti2005principled,itti2009bayesian,baldi2010bits}. 

Bayesian Surprise~\cite{itti2005principled,itti2009bayesian,baldi2010bits} is defined as a measure of divergence between prior and posterior distributions. In other words, it captures surprise as the number of bits in this divergence. If prior and posterior are close, we are not surprised after observing the dataset $\mathcal{D}$.

%One problem when computing surprise is that for most prior distributions, no closed analytical form exists for the posterior. This arises due to the complex integral in the denominator. Nevertheless, a common approach when using Bayesian inference is to choose {\it conjugate priors}.  These are priors that guarantee that both prior $p(\theta)$ and the posterior $p(\theta \mid \mathcal{D})$ are captured by the same pdf.

Our focus will be on measuring surprise based on a node's relative position in a ranking. %To do so, we employ the Beta distribution as a prior. 
Now, consider a ranking such as $a_3, a_1, a_2$. Here, some artist $a_3$ was ranked above all others, and $a_2$ was ranked above $a_1$. The relative position of $a_3$ is 1, as all other nodes are below it. $a_1$ has a relative position of $\frac{2}{3}$ and $a_2$ has a relative position of $\frac{1}{3}$.

Our dataset $\mathcal{D}$ will thus be the set: $\{x_1 = \frac{2}{3}, x_2 = \frac{3}{3}, x_3 = \frac{1}{3}\}$. In other words, $x_i = \frac{g_{t,i}}{|\mathcal{V}_t|}$, where $g_{t,i}$ is the set of nodes with a score equal to or greater than node $i$ at time $t$, and $|\mathcal{V}_t|$ is the number of nodes at time $t$.
Being simple fractions in the $[0, 1]$ range, each one of these values may be modeled as a Bernoulli distribution, i.e., $x_i \in \{0, 1\}$ and $x_i \sim Bernoulli(\theta)$. To measure surprise, we can employ the Beta distribution as a prior as it is well known that the Beta distribution is the conjugate prior to the Bernoulli. %Here, $\theta$ is a single value capturing the fraction of values where $x_i = 1$. %This may be computed by the mean of the dataset $\theta = \{\hat{x}\}$. %Notice that what we observe an intuitive measure (fraction of positives). This measure is the one that maximizes the likelihood. % Now that we have the likelihood, a Bernoulli, and it's maximal parameter, $\theta$, we need a prior. 

Thus, assume a prior such that $\theta \sim Beta(\alpha_\theta, \beta_\theta)$, with $p(\theta)$ being the density function. The prior has parameters, denominated as hyper-parameters $\alpha_\theta$ and $\beta_\theta$, representing the hypothesis we use for computing surprise. Being the Beta the conjugate prior, the posterior is proven to be $\theta \mid \mathcal{D} \sim Beta(\alpha_{\theta\mid\mathcal{D}}, \beta_{\theta\mid\mathcal{D}})$, with $p(\theta \mid \mathcal{D})$ being its density. Here, $\alpha_{\theta\mid\mathcal{D}} = \alpha_\theta + g_{t,i}$. Also, $\beta_{\theta\mid\mathcal{D}} = \beta_\theta + (|\mathcal{V}_t| - g_{t,i})$. %$\mathbb{I}(\cdot)$ is an indicator function returning one whenever its input is true.

% Also notice from this example how the posterior parameters are in-fact a function of the data. $\alpha_{\theta\mid\mathcal{D}}$ is proportional to the number of positives, and $\alpha_{\theta\mid\mathcal{D}}$ to the number of negative. From the sum, it is easy to see that with large enough samples the choice of priors becomes irrelevant~\cite{gelman2013bayesian}.

%\subsubsection{Bayesian Surprise} \label{subsub:musicsup}

Bayesian Surprise~\cite{itti2005principled,itti2009bayesian,baldi2010bits} thus is computed as the divergence between prior and posterior distributions. 
A common choice is to use the Kullback-Leibler Divergence:
\begin{align}
    D_{KL}(p(\theta|\mathcal{D}) \mid\mid p(\theta)) = \int_\theta p(\theta|\mathcal{D}) \log_2 ({p(\theta|\mathcal{D}) \over p(\theta)}) d\theta
\end{align}
Due to the use of conjugate priors, closed-form solutions for the above measure already exists~\cite {baldi2010bits,penny2001dkldirichlet,soch2016kullback}. In particular, ours is the divergence of two Beta distributions~\cite{penny2001dkldirichlet}.  %We note that other divergences between probability distributions could be employed to measure surprise. $D_{KL}$, however, is the most widely adopted.

Following the properties of the divergence, surprise will always be positive. Lower values indicate less surprise, as we have better priors (or hypotheses) about the data. It equals zero in the unrealistic setting with a prior perfectly capturing the posterior. This is unrealistic, as the posterior itself transforms the prior after observing some data.

Now, the above definition is valid for a single choice of prior, or hypothesis, as we now call it. We can also consider a set of hypotheses: $\mathcal{H} = \{\theta_h\}$. Each hypothesis is a prior distribution: $\theta_h \sim Prior(\cdot)$. In the case of the Beta distribution prior, our set will comprise different choices for the hyper-parameters: $\alpha_{h,\theta}$ and $\beta_{h,\theta}$. That is, $\theta_h \sim Beta(\alpha_{h,\theta}, \beta_{h,\theta})$. %The subscript $h$ indicates one particular hypothesis.
Given that divergence is additive in nature, the overall surprise for various hypotheses is:
\begin{align}
    Sup(\mathcal{H}, \mathcal{D}) = \sum_{h=1}^{|\mathcal{H}|} D_{KL}(p(\theta_h|\mathcal{D}) \mid\mid p(\theta_h)) 
\end{align}
% \noindent Here, we imply that each $\theta_h$ is a different set of parameters for the same probability distribution.% define by the density function $p$ (dropped from the notation for clarity).

We currently have the required definitions for computing surprise on music influence networks. Nevertheless, we point out that these definitions are general enough to be adapted to different domains of the musical information retrieval literature. 
As stated, we shall compute surprise via {\it rankings} on the network. That is, each artist is ranked according to some centrality score. Each hypothesis is a prior distribution of the position where we expect the node's relative to be in the ranking. This expected position comes from previous snapshots of the network as we now detail.

In this paper, we consider two hypotheses. The first is called {\it Past Rank}. In this hypothesis, we define that for each node i: $\alpha_{i,\mathcal{\theta}} = g_{i,t-1}$ and that $\beta_{i,\mathcal{\theta}} = |\mathcal{V}_{t-1}| - \alpha_{i,\mathcal{\theta}}$. Here, the subscript indicates the hypothesis for node $i$ on timestamp $t$ as a prior over parameters $\theta$. With these hyperparameters, the expected value of the Beta distribution is equal to the relative position of the node on time $t-1$. This comes from the fact that the expected value of the Beta is: $\frac{\alpha_{i,\mathcal{\theta}}}{\alpha_{i,\mathcal{\theta}} + \beta_{i,\mathcal{\theta}}}$ Thus, Past Rank assumes that the node's position will not change from one snapshot to the next. 

The second hypothesis is called {\it Regular Growth}. Here, we state that: $\alpha_{\mathcal{\theta}} =  \frac{g_{i, t-1}^2}{g_{i, t-2}}$. And again, $\beta_{\mathcal{\theta}} = |\mathcal{V}_{t-1}| - \alpha_{\mathcal{\theta}}$. Here, we aim at capturing regularities in changes. To understand it, observe that $\frac{g_{i, t-1}}{g_{i, t-2}}$ captures the rate of change considering the last two snapshots. Consider, for example, that this rate of change is equal to two. When we multiply the rate by $g_{i, t-1}$, we state that we expect the node to rise two places. When the rate is below one, the node will fall in ranking. Thus, this hypothesis places the expected value of the Beta distribution in the position where we expect the node to be given the past two timestamps of the network.

%Some care needs to be taken when computing surprise based on these hypotheses using real-world data. The {\it Past Rank} surprise is only viable when an artist appears in at least two snapshots (the current and a previous one). {\it Regular Growth} needs at least three. %Also, it is naturally expected that depending on time between snapshots, the surprises observed may be more abrupt. % (e.g., one artist is already highly influential on the initial snapshot). %Thus, results need to be interpreted considering the dataset granularity.

\section{Results} \label{sec:dataset}

\begin{table}
\footnotesize
\begin{center}
\begin{tabular}{lrrrrr}
\toprule
 & Weighted & \# Artists & \# Edges & Range & $\Delta$ \\
\midrule
\rowcolor{gray!6}  AllMusic & False & 32,568 & 119,961 & 1940-2019 & 10 \\
WhoSampled & True & 166,016 & 603,487 & 1940-2019 & 1\\
\bottomrule
\end{tabular}
\end{center}\vspace{-1em}
 \caption{Overview of the dataset. $\Delta$ refers to the granularity in years. When $\Delta=10$, each snapshot is a decade.}\vspace{-1em}
 \label{tab:data}
\end{table}

Our results focus on influence networks. The AllMusic network is publicly available (see~\cite{figueiredo2019disruption})\footnote{\url{https://github.com/flaviovdf/allmusic-disruption}}. Using a large seed of 73,000 thousand AllMusic URLs present in MusicBrainz\footnote{\url{https://musicbrainz.org/}}, AllMusic's influence network was captured via snowball sampling. After filtering nodes with at least one edge, the network comprises of 32,568 artists connected by 119,961 edges. Edges capture that one artist influenced another and were defined by the editors of the website. In this dataset, each snapshot comprises a decade.% (shown as $\Delta$ on the table).

The WhoSampled dataset was provided after an agreement with {\it WhoSampled.com}. WhoSampled lists for several songs the other songs that either sampled, remixed or covered it. The provided dataset contained a total 1,250,246 songs. For our study, we aggregate from the song level to the artist level. That is, we define an edge between artists that sampled, remixed or covered other artists. The weight of this edge is the total number of interactions (samples + remixes + covers). After filtering the data to consider the same temporal range as AllMusic, we are left with 166,016 nodes and 603,487 edges. %This dataset enables for a yearly analysis. 

% \section{Results} \label{sec:results}

% We now apply our surprise models to our datasets. First, we compare Bayesian Surprise (Section~\ref{subsec:face}). Next, we validate how it can be used to profile the dynamic of individual artist trajectories in (Section~\ref{subsec:trajectories}).

% \subsection{Comparison with Other Approaches} \label{subsec:face}

% We compare Bayesian Surprise with different correlation measures for rankings. These include classical correlations, such as Spearman~\cite{spearman1904coeff} and Kendall~\cite{kendall1938new} measures, as well the recent Weighted Kendall~\cite{vigna2015weighted} measure. For this latter, weights are centrality scores: $s_{t,n_i} = score(\mathcal{G}_{t}, n_i)$.

\begin{table}
\footnotesize
\begin{center}
\begin{tabular}{lrr}
\toprule
 & Kendall & Spearman \\
\midrule
\rowcolor{gray!6}  AllMusic & 0.026  & 0.036 \\
WhoSampled & 0.203 & 0.290\\
\bottomrule
\end{tabular}
\end{center}\vspace{-1em}
 \caption{Correlation coefficients between PageRank and Disruption on the Last Snapshot of the two datasets.}\vspace{-1em}
 \label{tab:corr}
\end{table}

\subsection{Motivating Surprise}

With these datasets, we present some motivating examples of Bayesian Surprise. Our goal here is to show that a single ranking does not tell the whole story of an artist. By combining several rankings in a principled manner, Bayesian Surprise mitigates this issue.

Thus, our first results show that Pagerank and Disruption play different roles when ranking nodes. In Table~\ref{tab:corr}, we present both Spearman~\cite{spearman1904coeff} and Kendall~\cite{kendall1938new} correlations of the two centrality scores. Scores were measured on the last snapshot for each dattaset. Correlations for both datasets are low, thus showing evidence that Pagerank and Disruption scores capture complementary aspects of the network topology. Such a result motivates the need to combine both scores. Bayesian Surprise provides a principled way to achieve this goal. Next, we further explore individual trajectories.% one of the primary motivations of our work.

% \begin{figure*}[t!]
%  \centering
% \subfloat[Spearman]{\includegraphics[width=0.32\linewidth]{figs/kendall.pdf}}\hfill
% \subfloat[Kendall]{\includegraphics[width=0.32\linewidth]{figs/kendall.pdf}}\hfill
% \subfloat[Weighted Kendall]{\includegraphics[width=0.32\linewidth]{figs/wkendall.pdf}}
% \caption{Correlation from one snapshot to the next.}~\label{fig:corr}\vspace{-1em}
% \end{figure*}

\begin{figure}[t!]
\centering
\subfloat[Pagerank]{\includegraphics[width=0.75\columnwidth]{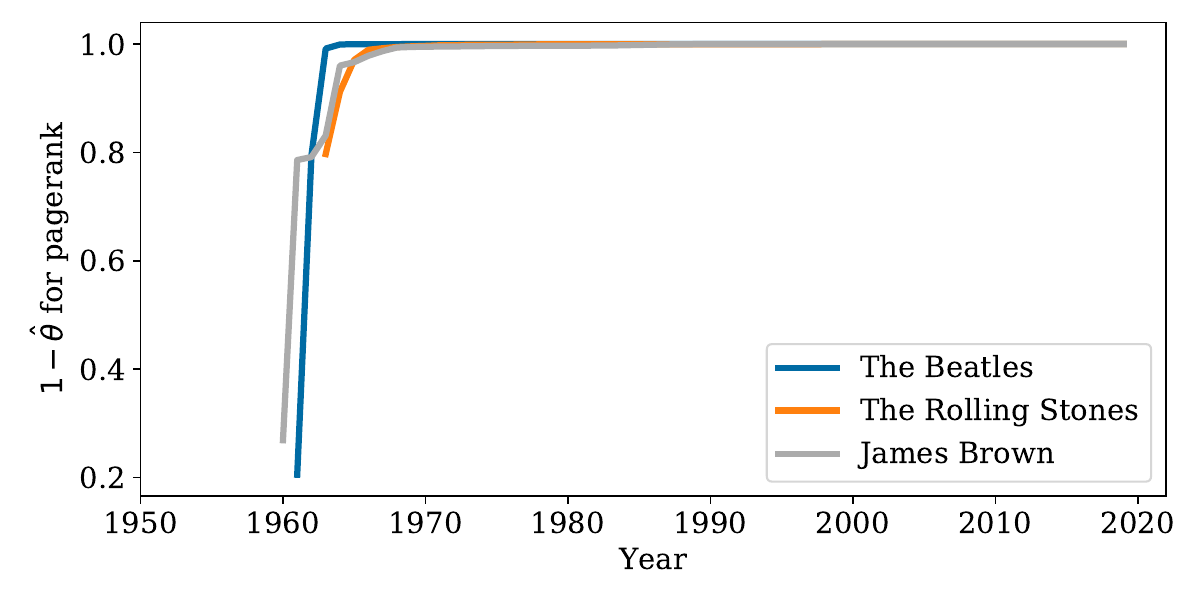}}\\\vspace{-1em}
\subfloat[Disruption]{\includegraphics[width=0.75\columnwidth]{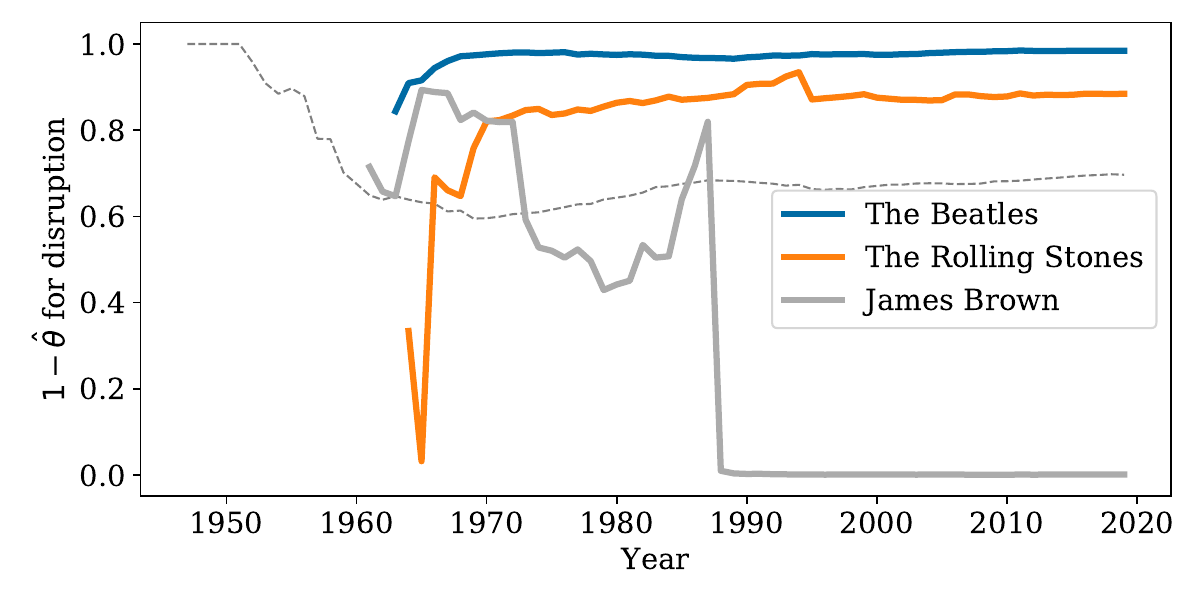}}
\caption{Comparing the position in rankings. The dashed line indicates the position where Disruption changes sign.}~\label{fig:top}\vspace{-1em}
\end{figure}

% We now argue that simply measuring correlations across snapshots does not lead to useful insights on network evolution. Thus, in Figure~\ref{fig:corr} we show correlation scores from $\mathcal{G}_{t-1}$ to $\mathcal{G}_{t}$. In other words, how correlated are rankings from one snapshot to the next. From the figure, we can see that for the WhoSampled dataset, regardless of the choice of correlation, the measures quickly rise to high correlations and remain roughly constant. For AllMusic, measures tend to show a roughly constant increase. 

% These results indicate that looking and how the network changes using correlations only will not indicate snapshots where the topology has been significantly altered. Also, these scores do not translate to the node level, something we explore next with some face validity examples.

In Figure~\ref{fig:top}, we show three artists their position in the ranking for WhoSampled. These are three of the most influential artists both scores. Positions are shown as $1-\theta$, thus higher values indicate higher positions. The figure shows that The Beatles quickly rises to influence in terms of Pagerank (Figure~\ref{fig:top}-a). When we consider Disruption (Figure~\ref{fig:top}-b), the band takes some time for the band to rise to the top ranks of the scores. The Rolling Stones follows a similar pattern, but with more ``jumps''.

It is also interesting how James Brown both rises and falls in Disruption over time. Eventually, the artist declines in Disruption but remains stable in Pagerank. Looking at only one of the scores would only tell part of this story. Bayesian Surprise can highlight the changes for a single artist, such as James Brown's trajectory, by adding several hypotheses. We further explore this next. %but it is also able to compare artists (they are part of the same rank). 

\begin{figure}[t!]
 \centerline{
 \includegraphics[width=.75\columnwidth]{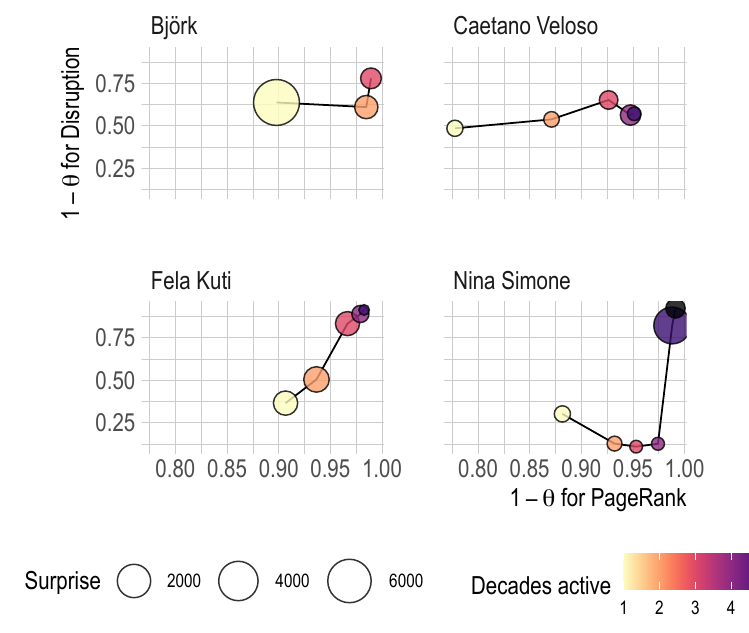}}

 \caption{Trajectories of four artists in AllMusic as a function of their Pagerank, Disruption and surprise.}\vspace{-1em}
 \label{fig:am-trajs}

\end{figure}

\begin{figure}[t!]
 \centerline{
 \includegraphics[width=.72\columnwidth]{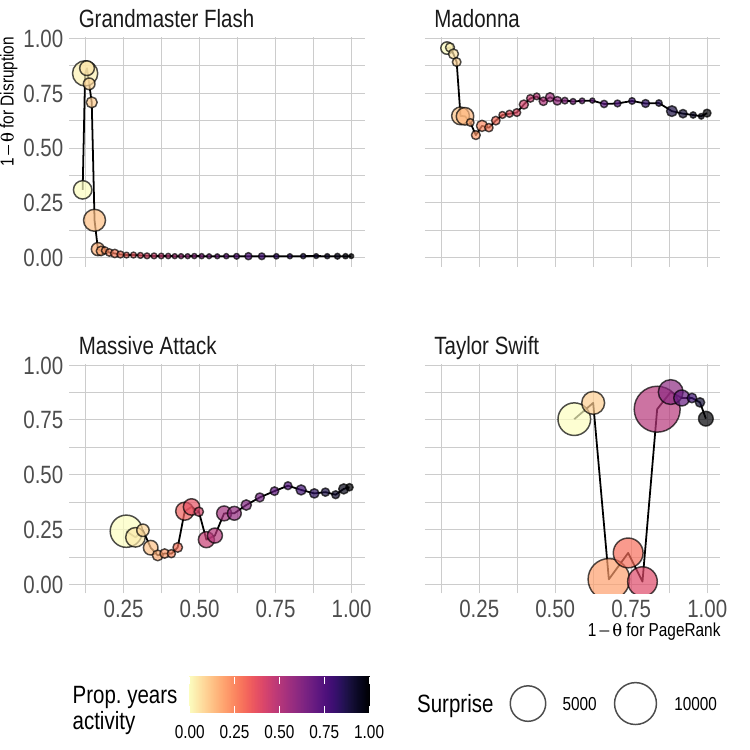}}
 \caption{Trajectories of four artists in WhoSampled.}\vspace{-1em}
 \label{fig:ws-trajs}

\end{figure}

\subsection{Making use of Surprise}

Next, in Figures~\ref{fig:am-trajs} and~\ref{fig:ws-trajs}, we show the trajectories of four artists in each dataset as connected scatter plots. Here, the surprises of the two hypotheses were summed up. To provide a concise view, we also sum the hypotheses in the PageRank and Disruption section. The x-axis shows the position in Pagerank, while the y-axis shows the position in Disruption. Surprise is shown by the size of each point. 

Starting from Figure~\ref{fig:am-trajs}, some artists such as Bjork are surprising early in their careers (notice that larger points are the brighter ones). That is, the artist quickly rises to top positions in each centrality score. Further movements from the artist are less surprising, but still highlighted (when we compare with the size of other artists) as Bjork continues to rise to the top of PageRank and Disruption rankings. Artists like Caetano Veloso have smoother trajectories, being thus less surprising. To understand this, recall both of our hypotheses are functions of the previous snapshots. Hypothesis {\it Reg. Growth}, in particular, aims at capturing these smoother transitions. Nina Simone shows an example of abrupt transitions being highlighted due to a surprisingly higher disruption later in her career.

In WhoSampled (Fig.~\ref{fig:ws-trajs}), we are able to observe more granular trajectories. The Hip Hop pioneer Grandmaster Flash provides an example of an initially highly disruptive career that afterwards gains increasing relevance, but acts as a consolidator. Additionally, his peak and sudden decrease of disruption are surprising, marking rapid creation of a large and consolidated stream of artists. Massive Attack has a somewhat similar trajectory, with most surprise in its early years, but maintaining some disruption, likely being sampled by artists that do not sample its influences. Madonna and Taylor Swift are examples of artists whose trajectories have peak surprise points after their initial years in the dataset. Finally, for these examples changes in Disruption are more highlighted than changes in Pagerank. Given that these are popular artists, we can notice that their evolution in Pagerank scores (x-axis) is quite smooth over time. Thus, they are less surprising. 

This is most evident with Taylor Swift, whose trajectory has multiple surprising changes in disruption, denoting points where there is first a wave of artists who are influenced by her and by her influences (drop in disruption), and afterward, a wave of artists who are influenced by her but do not share her influences (increase in disruption on 6th point in her trajectory). When we look into Taylor Swift's disruption values, they are reduced tenfold from 2011 to 2013 (from 0.1 to 0.01). Later in the dataset, the artist regains her position.

Our previous results motivate Bayesian Surprise over alternatives. The trajectories we presented here exemplify how Surprise can guide the analysis in identifying relevant phenomena in the evolution of dynamic networks. %We note the same approach could be used to highlight surprising phenomena in larger dimensional spaces (more centralities and hypotheses).

\section{Conclusions}

In this paper, we derived a Bayesian Surprise measure based on rankings that was validated on real-world datasets. Our is the first work to provide this measure for social networks and for the music domain. Secondly, our analysis on the temporal nature (via artist trajectories) of music influence networks provider novel insights into how influence evolves based on two centrality scores. In particular, our results help understand music datasets from a historical perspective.

%we present the first application of Bayesian Surprise to musical influence networks. Even though we apply surprise to rankings in musical influence networks only, the ideas presented here are suitable for other musical information retrieval areas. More importantly, our results help understand music datasets from a historical perspective.% (such as AllMusic) and change (our yearly analysis in WhoSampled).

%As future work, we note that with the correct choice of priors, Bayesian Surprise is also suitable to understand different music datasets such as MFCCs (considering MFCCs as images, this is the approach of~\cite{itti2005principled}). 

% In this sense, we view two main contributions from our work:

%In this sense, we view three main contributions from our work. Firstly, we derive a Bayesian surprisal measure based on rankings which was validated on real world datasets. Secondly. not only is the Bayesian surprisal framework useful for for networks, the framework may be ported to other domains of MIR. Lastly, our analysis on the temporal nature (via artist trajectories) of music influence networks provider novel insights into how influence evolves based on two centrality scores. % Pagerank and Disruption.

\balance
\bibliography{ISMIRtemplate}

\end{document}